# Topological Semimetal-Insulator Quantum Phase Transition in Zintl Compounds Ba$_2$X (X=Si, Ge)

Ziming Zhu[†§], Mingda Li[‡], Ju Li[†§]*

[†]Frontier Institute of Science and Technology, and State Key Laboratory for Mechanical Behavior of Materials, Xi'an Jiaotong University, 710054, Xi'an, People's Republic of China

[‡]Department of Mechanical Engineering, Massachusetts Institute of Technology, Cambridge, MA 02139, USA

[§]Department of Nuclear Science and Engineering and Department of Materials Science and Engineering, Massachusetts Institute of Technology, Cambridge, Massachusetts 02139, United States

*Correspondence and requests for materials should be addressed J.L. (email: liju@mit.edu).

**Abstract:** By first-principles calculations, we find that Ba$_2$X (X=Si, Ge) hosts a topological semimetal phase with one nodal ring in the $k_x = 0$ plane, which is protected by the glide mirror symmetry when spin-orbit coupling (SOC) is ignored. The corresponding drumheadlike surface flat band appears on the (100) surface in surface Green's function calculation. Furthermore, a topological-semimetal-to-insulator transition (TSMIT) is found. The nodal line semimetal would evolve into topological insulator as SOC is turned on. The topologically protected metallic surface states emerge around the $\bar{\Gamma}$ point, which could be tuned into topologically-trivial insulator state by more than 3% hydrostatic strain. These results reveal a new category of materials showing quantum phase transition between topological semimetal and insulator, and tunability through elastic strain engineering.

## I. Introduction

Topological insulator (TI) [1, 2] possesses exotic metallic surface states but gapped bulk states. It has attracted enormous attention since the theoretical proposal of 2D TI in HgTe/CdTe quantum well and 3D TIs in $Bi_2Se_3$ family of compounds [3-5], and experimental verifications thereafter [1, 2, 6]. The most striking characteristics of TI is the robust edge/surface state in 2D/3D system, sheltered against backscattering from non-magnetic impurities as long as bulk gap remains open and the time reversal symmetry is preserved [1-8]. Recently, it has been shown that the band topology could be extended into semi-metallic systems where some band touches are exactly located at the Fermi level [9-16]. Based on the different band touches in the topological semimetal (TSM), Dirac point, nodal line or ring would form on the Fermi surface as a consequence of nontrivial band crossing guaranteed by crystal symmetry and topology[11, 12, 17-23]. For instance, starting with two theoretical proposals in $Na_3Bi$ and $Cd_2As_3$ [11, 12], TSM phase has been experimentally confirmed as an analogue of 3D graphene for band structure with linear dispersion at different crystal planes [24, 25]. In particular, Fermi arcs are expected to be detectable spectroscopically in TSM with the breaking of either time-reversal symmetry or inversion symmetry (Weyl semimetal), which have been verified in non-centrosymmetric TaAs-class materials [13, 16, 26, 27].

3D TSM could be driven into 3D TI with the breaking of rotational symmetry. For example, Zeng *et al.* found through first-principles calculations that rare earth monopnictide LaN is a TSM possessing nodal ring when SOC is neglected, but exhibiting TI behavior when SOC is included [28]. Moreover, a nodal ring could be turned into nodal point by SOC in the absence of inversion symmetry, dubbed Weyl point [13, 16]. The common feature between them is that the nodal ring from the crossing between the conduction band and the valence band is protected by the reflection symmetry and the time-reversal symmetry, and SOC plays significant role in driving the quantum phase transition from TSM into TI. However, there is a lack of understanding on how SOC affects the nodal line state respecting the glide mirror symmetry. In this article, based on first-principles calculations, we find a TSM phase in Ba$_2$X (X=Si, Ge) with a nodal ring in the $k_x = 0$ plane, which would evolve into TI phase by SOC. Also, such TI could be further tuned into normal semiconductor by hydrostatic strain (HSS). These investigations provide the evidences to understand the impact of SOC on the nodal line semimetal and strain-induced topological-semimetal-to-insulator transitions (TSMIT), and may also offer an avenue to searching for new topological materials and quantum phase transitions.

## II. Crystal structure and Methodology

Ba$_2$X (X=Si, Ge) belongs to the orthorhombic crystal system with the space group Pnma (No. 62). In this study, for brevity we take Ba$_2$Si as an example to introduce the electronic structures, as the similarity with Ba$_2$Ge is strong. The lattice constants are *a*=5.418 Å, *b*=8.524 Å and *c*=10.174 Å [29]. The corresponding primitive cell, shown in Fig. 1(a), consists of eight Ba and four Si. Ba contains two inequivalent 4c sites, Ba$_1$ (0.0185, 0.25, 0.6756) and Ba$_2$ (0.1513, 0.25, 0.083), whereas Si occupies the 4c site (0.251, 0.25, 0.4016). The bulk Brillouin zone (BZ) is illustrated in Fig. 1(b) along with the (100) BZ surface. Vienna ab initio simulation package (VASP) was employed to calculate bulk band structure with the generalized gradient approximation (GGA) density functional and the projector augmented wave (PAW) method [30-32]. A Monkhorst-Pack **k**-mesh of 11×11×11 was used for the bulk calculation to sample the Brillouin zone and the energy cutoff for plane wave basis was set to 560 eV [33]. For convergence of electronic self-consistent calculations, a total energy difference criterion was defined as $10^{-8}$ eV. All atoms were fully relaxed in the internal coordinates until the total residual force was less than $10^{-3}$ eV/Å. To calculate the surface state, $d$ component of Ba and $p$ of Si were obtained as the atomic-like Wannier functions for tight-binding model as in Ref. [34]. The phonon spectrum was calculated using a supercell approach within the PHONON code [35].

### III. Results and Discussions

In the absence of SOC, band structure demonstrates the feature of semimetal phase in bulk $Ba_2Si$ as depicted in Fig. 1(c), where a band inversion happens along the $Y - \Gamma - Z$ direction, resulting in nodal ring from the crossing between the valence band and the conduction band that stays in the $k_x = 0$ plane (the inset of Fig. 1(c)). The glide symmetry $G_x = (M_x | (\frac{1}{2}, 0, 0))$ is the combination of mirror reflection symmetry $M_x : (x, y, z) \rightarrow (-x, y, z)$ and a translation by a half lattice vector $\tau_x = (\frac{1}{2}, 0, 0)$. Thus, the eigenvalue of $G_x$ is the product of the eigenvalue of $M_x$ and the phase factor induced by $\tau_x$. The irreducible representations of the little group of **k** point in the $k_y - k_z$ plane are determined by the eigenvalue of $M_x$. As the Hamiltonian without SOC is spin-rotation invariant, i.e. $M_x^2 = 1$, the eigenvalue of $M_x$ is +1 or -1 (Fig. 1 (c) [36]. According to this argument, two bands with different representations will not induce a band gap when they cross each other. Therefore, the appearance of the nodal ring in the absence of SOC is protected by the glide mirror symmetry. Taking SOC into consideration (Fig. 1 (d)), two bands move away from each other and it is fully gapped along the whole high symmetry direction. The opened gap is about 25 meV. This is identified as TSM-TI quantum phase transition (see below). From the comparisons of electronic structures with and without SOC, the variation of whole band structure is rather small because the elements involved have light *Z*. It is worth mentioning that an inverted band order state already exists in TSM phase as compared to the normal semiconductor phase, such as $Sr_2Pb$ and $Bi_2Te_3$ [5, 37] when SOC is turned off, although they would evolve into TI phase in the presence of SOC. Additionally, the stability of Zintl compound $Ba_2Si$ has been checked by computing the phonon dispersion as illustrated in Fig. 2, with no imaginary frequency along the high symmetry line.

Figs. 3 (a)-(c) show the band-orbital characteristics around the $\Gamma$ point to clarify the topological property, which has been analyzed from the wavefunctions of first-principles calculations. In the case of strain-free crystal, the valence band at the $\Gamma$ point (VBG) is mainly composed of Ba(*d*) orbital, while the conduction band at the $\Gamma$ point (CBG) consists of Si(*p*) orbital no matter whether SOC is introduced. With 5% of applied HSS, VBG and CBG exchange their atomic orbital with opposite band order, which obviously belongs to topologically trivial semiconductor. Fig. 3 (d) illustrates that the energy difference between VBG and CBG decreases as a function of HSS, where topological phase transition occurs at about 3%. However, the energy gap between the valence band maximum and the conduction band minimum is almost unchanged as the strength of SOC does not rely much on strain. In reality, $Ba_2Si$ possesses *p-d* band inversion, comparing with *p-s* band inversion in $Sr_2Pb$ system though they share the similar crystal structure [37].

To confirm the topological nature of $Ba_2Si$ in the case of small but finite SOC as discussed above, we followed the parity criteria proposed by Fu and Kane [38] for such inversion symmetry system, i.e. calculating the product of the parity of Bloch wavefunction for the occupied band at the time reversal invariant momenta (TRIM) in the first Brillouin zone. There are seven TRIMs with '+' of the parity at X, Y, Z, S, U, T, and R, whereas the parity of the $\Gamma$ point depends on whether band inversion occurs as listed in Table I. For zero-strain case, the topological invariant $Z_2$ is 1. Furthermore, HSS effect could induce a transition from TI into topologically-trivial semiconductor, providing the possibility to study the tunability of topological phase in the $Ba_2Si$ system.

To further investigate the nontrivial surface state, we performed calculation of projected band structure in semi-infinite $Ba_2Si$ onto the (100) surface, by using iterative Green's function method [39] implemented in Wannier_tools package [40] from tight-binding model with Wannier function constructed from plane-wave solutions. Fig. 4 (a) illustrates that a drumheadlike flat band emerges around the Fermi level connecting two crossing point, which actually will form a drumheadlike Fermi surface. It should be pointed out that this type of flat band might be the key for high-temperature superconductor [41]. Taking SOC into account, one could see that there is a metallic surface state around the $\bar{\Gamma}$ point, coinciding with the previous calculation of $Z_2$. In $Ba_2Si$ the topologically protected surface state appears on the (100) surface, distinct from that of $Sr_2Pb$ on the (010) surface stemming from the distinct band inversion [37] structure. As SOC strength may be lowered or even diminished with lighter element, the surface density of states (SDOS) of nodal line may reappear by replacing Ba by Mg or Ca [22].

In summary, we report the quantum phase transition in $Ba_2Si$ from nodal line semimetal into topological insulator as a result of SOC based on first-principles calculation. The nodal line could still survive in the glide mirror symmetry system without including SOC, but would be fully gapped when SOC is introduced. Using Green's function calculation, we find that the nodal line semimetal demonstrates the drumhead flat band around the Fermi level, which turns into the metallic surface state with SOC. Our work also shows that such topological phase could totally disappear by applying more than 3% of HSS.

## Acknowledgements

We thank L. Fu, J.W. Liu and Q.S. Wu for helpful conversations. We would thank Q.S. Wu for helping calculate the eigenvalue of the glide mirror symmetry. This work is supported by National Basic Research Program of China, under Contract no. 2012CB619402, and a grant-in-aid of 985 Project from Xi'an Jiaotong University. J.L. acknowledges support by NSF DMR-1410636. M.L. would like to thank support by S3TEC, an EFRC funded by DOE BES under Award No. DE-SC0001299/DE-FG02-09ER46577.

## Competing financial interests

The authors declare no competing financial interests.

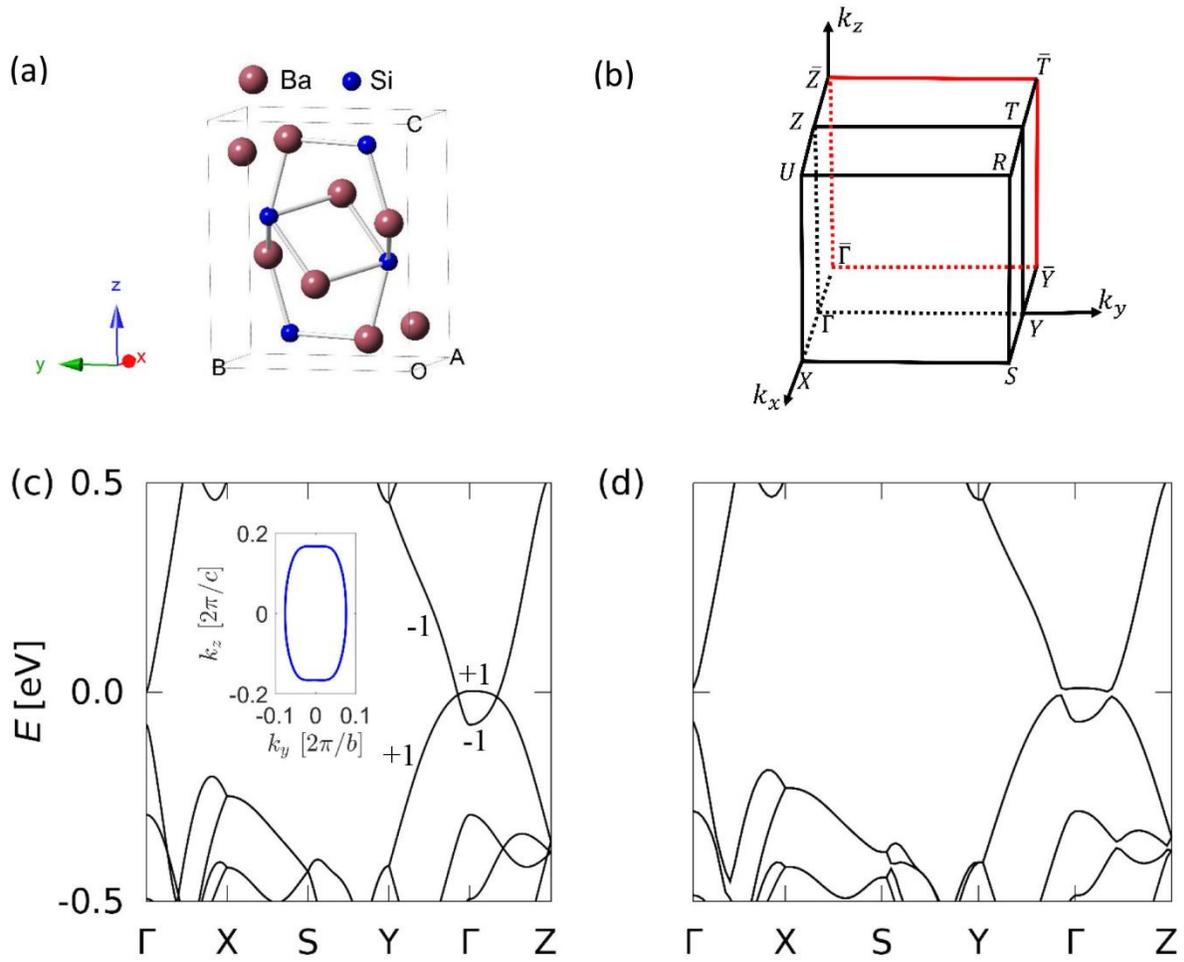

Figure 1. (a) Side view of primitive unit cell of Ba$_2$Si along the *x* axis. (b) Three dimensional bulk Brillouin zone and its projection onto the (100) surface, as well as high-symmetry points. Band structures from first-principles calculation (c) without and (d) with SOC. $E_F$ is set to be zero in all panels. The number +1 or -1 labels the eigenvalue of the glide mirror plane ($k_x = 0$). The inset shows the nodal ring within the momentum space of $k_x = 0$.

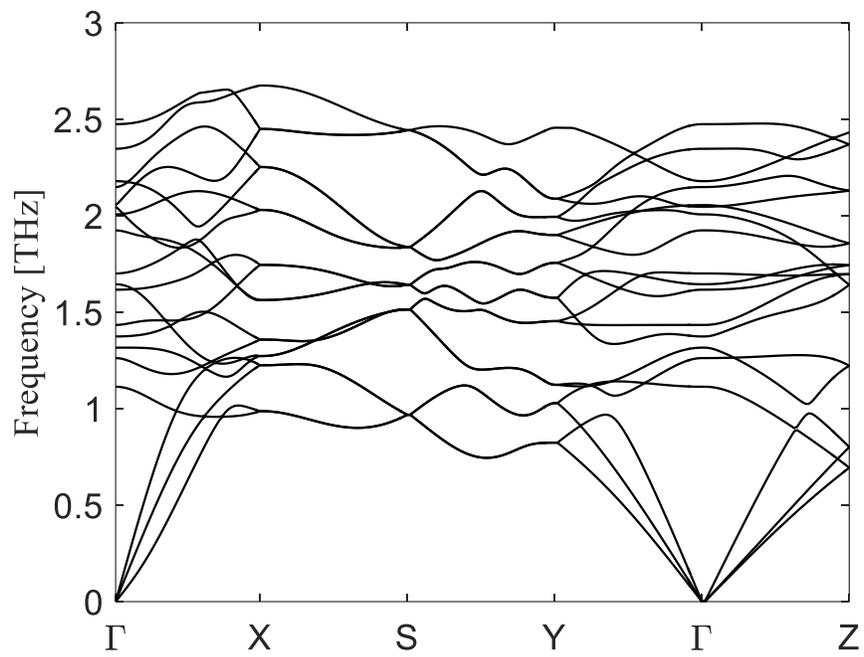

Figure 2. Phonon dispersion in Ba$_2$Si along the high-symmetry directions.

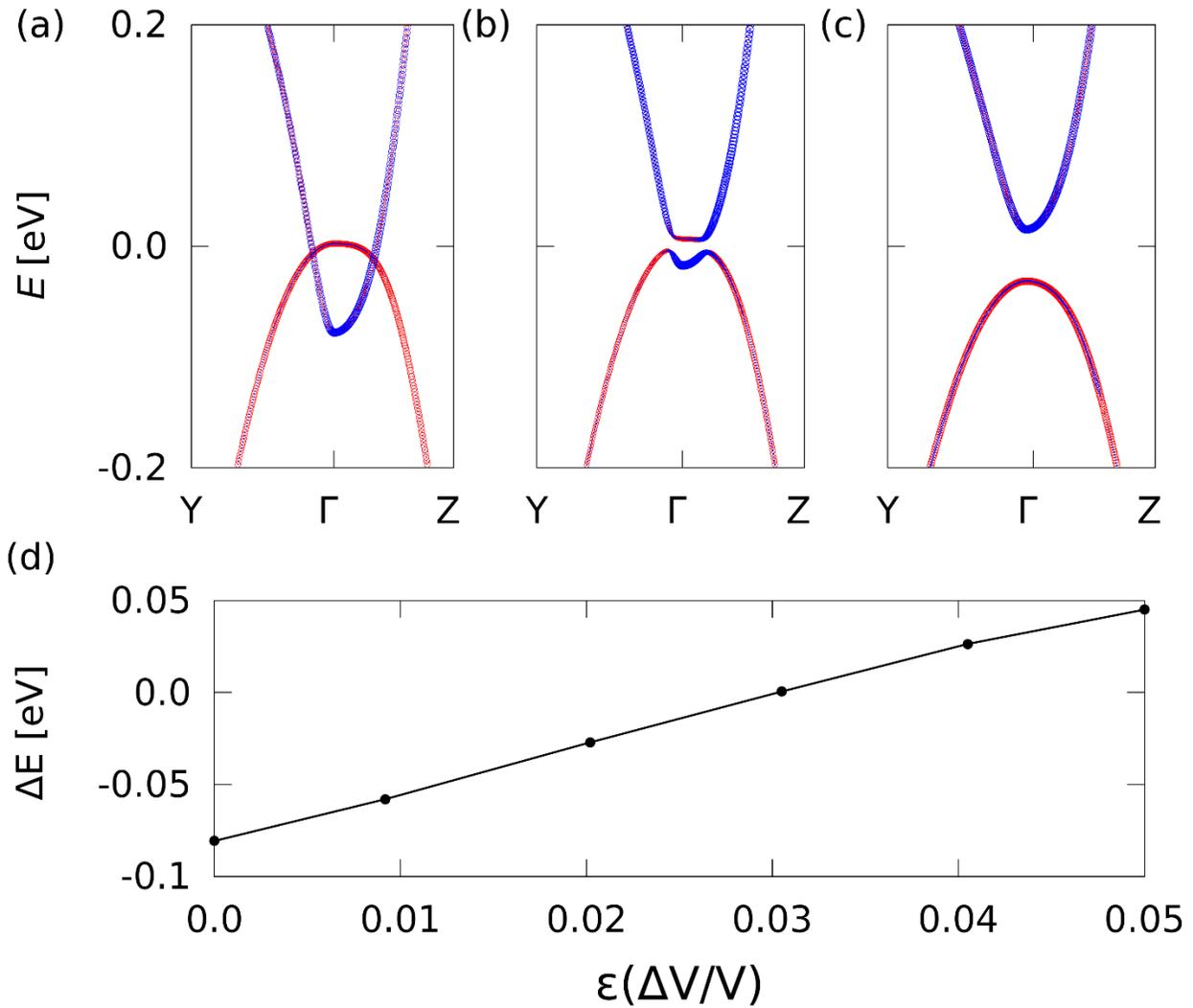

Figure 3. The comparisons of the component in different atomic orbitals along the $Y-\Gamma-Z$ direction among (a) without SOC, (b) ((c)) zero-strain (5% of HHS) with SOC, where the weight of atomic orbital Si($p$) (Ba($d$)) is proportional to the radius of blue (red) circle. Band inversion could be seen clearly around the $\Gamma$ point. (d) Energy difference between VBG and CBG (at the $\Gamma$ point) vs HSS, $\varepsilon(\Delta V/V)$.

|            | Γ | X | Y | Z | S | U | T | R | $Z_2$ |
|---|---|---|---|---|---|---|---|---|---|
| SOC, HSS=0% | − | + | + | + | + | + | + | + | 1 |
| SOC, HSS=5% | + | + | + | + | + | + | + | + | 0 |

Table 1. The parity of eight time-reversal invariant **k**-points with SOC as discussed in Fig. 3 and their corresponding topological invariant.

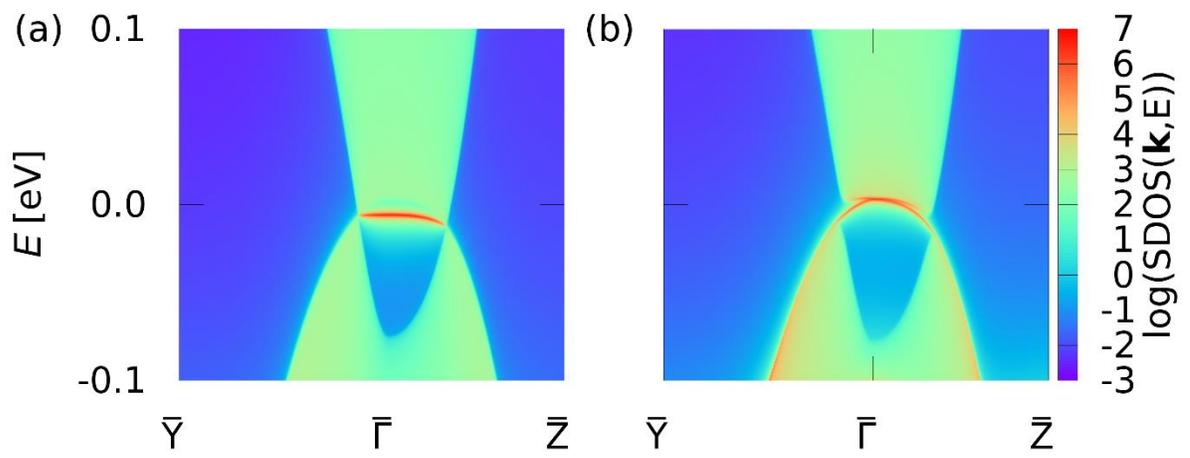

Figure 4. Surface density of states (SDOS) in $Ba_2Si$ (a) without and (b) with SOC. The red lines highlight the topologically protected metallic surface states along $\bar{Y}-\bar{\Gamma}-\bar{Z}$.